\documentclass{webofc}
\usepackage[varg]{txfonts}   % Web of Conferences font
\usepackage{hyperref}
\usepackage{url}
\hypersetup{colorlinks=true,citecolor=blue,urlcolor=blue,linkcolor=blue}
\begin{document}
\title{Quarkonium production in pp and heavy-ion collisions}

\author{\firstname{Taesoo} \lastname{Song}\inst{1}\fnsep\thanks{\email{t.song@gsi.de}} \and
        \firstname{Joerg} \lastname{Aichelin}\inst{2,3}%\fnsep\thanks{\email{Mail address for second
%             author if necessary}} 
\and
        \firstname{Jiaxing} \lastname{Zhao}\inst{4,5}%\fnsep\thanks{\email{Mail address for last
%             author if necessary}}
\and
        \firstname{Pol B.} \lastname{Gossiaux}\inst{2}%\fnsep\thanks{\email{Mail address for last
%             author if necessary}}
\and
        \firstname{Elena} \lastname{Bratkovskaya}\inst{1,4,5}%\fnsep\thanks{\email{Mail address for last
%             author if necessary}}
        % etc.
}

\institute{GSI Helmholtzzentrum f\"{u}r Schwerionenforschung GmbH, Planckstrasse 1, 64291 Darmstadt, Germany 
\and
SUBATECH UMR 6457 (IMT Atlantique, Universit\'{e} de Nantes,
	IN2P3/CNRS), 4 Rue Alfred Kastler, F-44307 Nantes, France
\and
Frankfurt Institute for Advanced Studies, Ruth-Moufang-Strasse 1, 60438 Frankfurt am Main, Germany
\and
Institute for Theoretical Physics, Johann Wolfgang Goethe Universit\"{a}t, Frankfurt am Main, Germany
\and
Helmholtz Research Academy Hessen for FAIR (HFHF),GSI Helmholtz Center for Heavy Ion Research. Campus Frankfurt, 60438 Frankfurt, Germany
          }

\abstract{We describe bottomonium production not only in pp collisions but also in heavy-ion collisions by using the Remler's formalism where quarkonium density operator is applied to all possible combination of heavy quark and heavy antiquark pairs.
In pp collisions heavy (anti)quark momentum is provided by the PYTHIA event generator after rescaling $p_T$ and rapidity to imitate the FONLL calculations.
Then spatial separation between heavy quark and heavy antiquark is introduced based on the uncertainty principle.
In heavy-ion collisions quarkonium wavefunction changes with temperature assuming heavy quark potential equals the free energy of heavy quark and heavy antiquark system in heat bath.
The density operator is updated whenever heavy quark or heavy antiquark scatters in QGP produced in heavy-ion collisions.
Our results are consistent with the experimental data from ALICE and CMS Collaborations assuming that the interaction rate of heavy (anti)quark in quarkonium is suppressed to 10 \% that of unbound heavy (anti)quark.
We also find that off-diagonal recombination of bottomonium barely happens even in Pb+Pb collisions at $\sqrt{s}=5.02$ TeV.
}
\maketitle
\section{Introduction}
\label{intro}
Ultra-relativistic heavy-ion collisions produce an extremely hot and dense matter which is supposed to cross over the phase boundary between hadron gas and quark-gluon plasma (QGP).
Quarkonium is a flavorless bound state of heavy quark and heavy antiqurk, whose suppression in heavy-ion collisions was suggested as a signature for QGP formation, because the binding of heavy quark pair is sensitive to the environment through the color screening and scattering with thermal partons~\cite{Matsui:1986dk}.
However, it was discovered that $J/\psi$ is less suppressed in mid-rapidity as well as in higher energy collisions, which indicates the regeneration of quarkonium from open heavy flavors~\cite{Manceau:2012ka}.
So the complete description of quarkonium production in heavy-ion collisions requires suppression as well as regeneration.

Quarkonium production takes two steps.
First, heavy quark pair is produced and then the produced heavy quark pair forms a bound state by emitting soft gluons to be a color singlet.
The first step is a hard process which can be described by pQCD, while the second step is a soft process which requires a model. 
In heavy-ion collisions, the first step is modified by so-called cold nuclear matter effects such as shadowing effects and the second step by hot nuclear matter effects.
In this study we try to describe quarkonium production not only in pp collisions but also in heavy-ion collisions by using the same Remler's formalism.

\section{Remler's formalism for quarkonium production}
\label{sec-remler}

The probability that a quarkonium eigenstate $i$ with momentum ${\bf P}$ is produced at ${\bf R}$ is given by~\cite{Remler:1975fm,Song:2023zma}
\begin{eqnarray}
(2\pi)^3\frac{d N_i}{ d^3R d^3P} =\sum \int \frac{ d^3r d^3p}{(2\pi)^3}\  \Phi^W_i({\bf r},{\bf p}) \prod_{j> 2} \int \frac{ d^3r_j d^3p_j}{(2\pi)^{3(N-2)}} W^{(N)}({\bf r_1,p_1,...,r_N,p_N}),
\label{eq:wig}
\end{eqnarray}
where ${\bf R}=({\bf r_1}+{\bf r_2})/2$, ${\bf r}={\bf r_1}-{\bf r_2}$, ${\bf P}={\bf p_1}-{\bf p_2}$, ${\bf p}=({\bf p_1}-{\bf p_2})/2$ with $\bf r_i$ and $\bf p_k$ being respectively the position and momentum of $k$'th (anti)heavy quark; .
$\Phi^W_i({\bf r},{\bf p})$ is the Wigner density of quarkonium wavefunction which is nonvanishing only for flavorless and colorless combination; $W^{(N)}({\bf r_1,p_1,r_2,p_2,...,r_N,p_N})$ is density matrix in Wigner representation of the $N$ (anti)heavy quarks produced in a proton-proton or a heavy-ion collision, which is simply taken for the classical phase space density distribution~\cite{Song:2023zma}.
%\begin{eqnarray}
%W^{(N)}\approx \prod_{i=1}^N h^{3N}\delta(r_i-r_i^*(t))\delta(p_i-p_i^*(t)),
%\label{deltas}
%\end{eqnarray}
The summation in the right hand side implies that all possible combinations of heavy quark pairs are considered to form quarkonium state $i$. 

In pp collisions the momentum of heavy quark and that of heavy antiquark are generated by the PYTHIA event generator followed by rescaling the transverse momentum and rapidity to imitate the FONLL calculations~\cite{Song:2015sfa}.
The spatial separation between heavy quark and heavy antiquark is given by the uncertainty principle~\cite{Song:2023zma}.
\begin{figure*}
\centering
\vspace*{1cm}       % Give the correct figure height in cm
\includegraphics[width=6.4cm,clip]{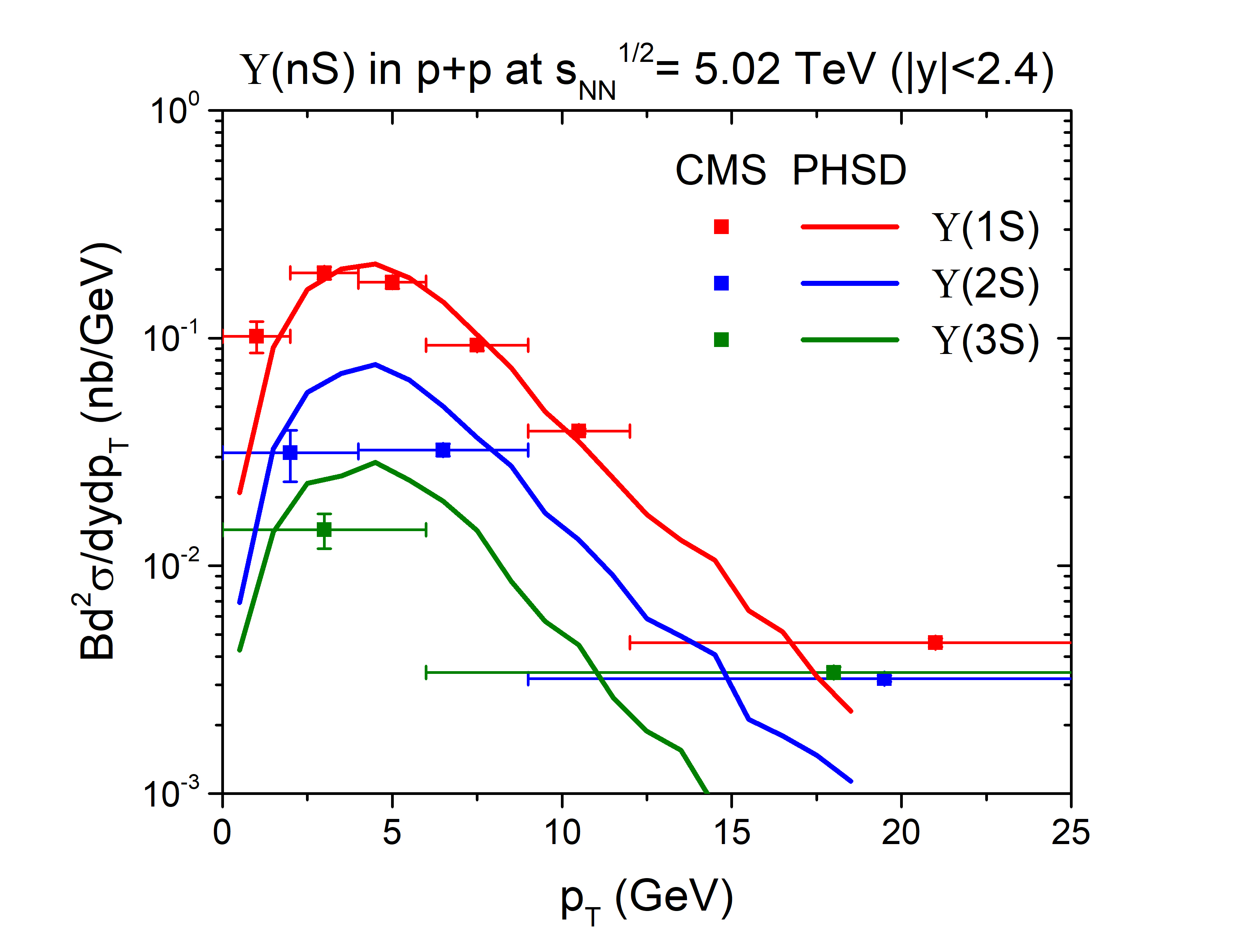}
\includegraphics[width=6.4cm,clip]{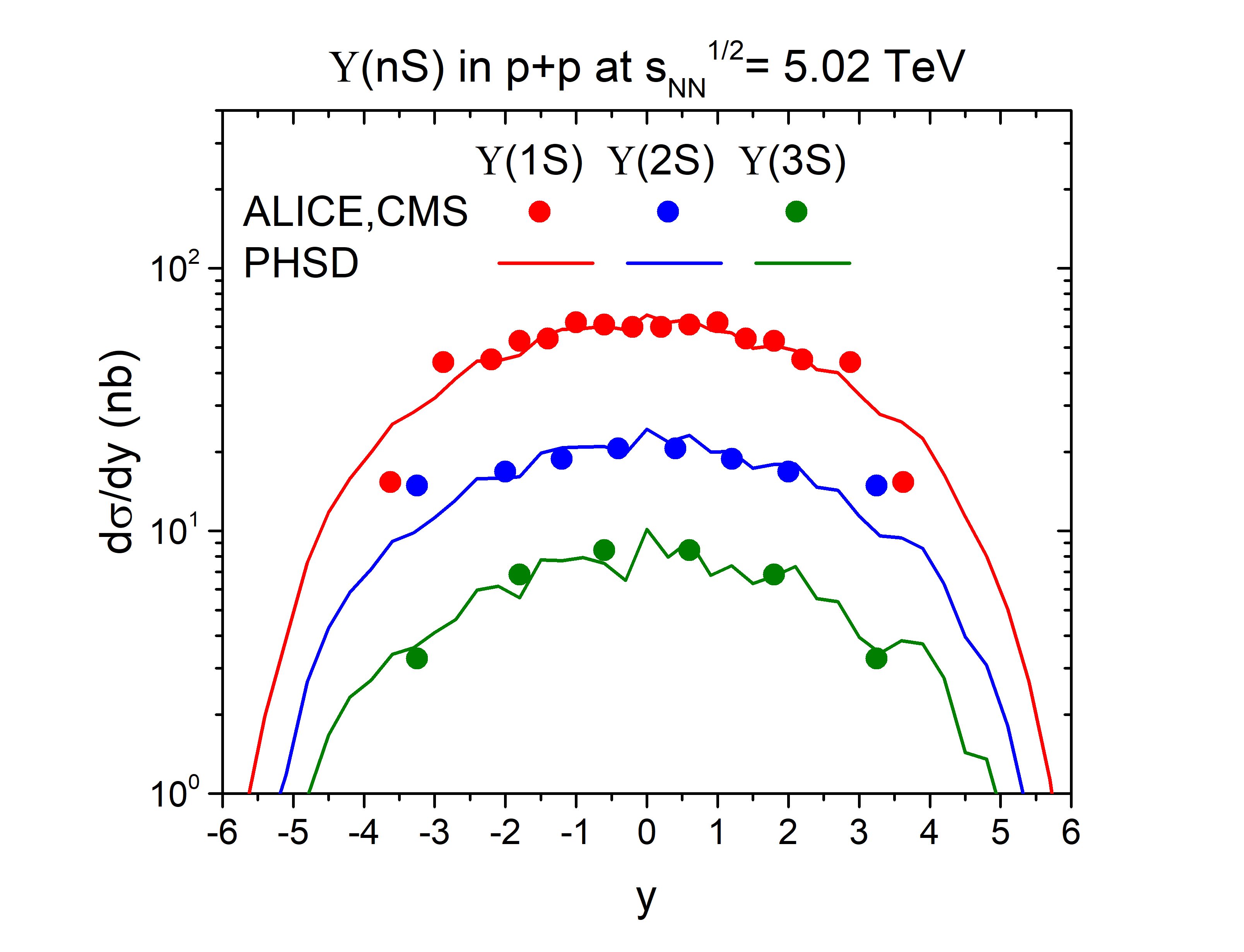}
\caption{(Left) $p_T$ spectra and (right) rapidity distributions of $\Upsilon (1S)$, $\Upsilon (2S)$ and $\Upsilon (3S)$ including the feed-out from excited states in pp collisions at $\sqrt{s}=5.02$ TeV, which are compared with experimental data from the CMS and ALICE Collaborations~\cite{CMS:2018zza,ALICE:2021qlw}.}
\label{pp}       % Give a unique label
\end{figure*}

Fig.~\ref{pp} shows the $p_T$ spectra and rapidity distributions of $\Upsilon (1S)$, $\Upsilon (2S)$ and $\Upsilon (3S)$ including the feed-out from excited states in pp collisions at $\sqrt{s}=5.02$ TeV.
One can see that the results are consistent with the experimental data from CMS and ALICE Collaborations~\cite{CMS:2018zza,ALICE:2021qlw}

A hot dense matter produced in relativistic heavy-ion collisions delays the formation of quarkonium.
We assume that the free energy of heavy quark pair is the heavy quark potential~\cite{Kaczmarek:2003ph}. 
In this case the dissociation temperature of $\Upsilon (1S)$ is roughly 3 $T_c$, while those of excited states are near $T_c$~\cite{Song:2023zma}.
The first Wigner projection of Eq.~(\ref{eq:wig}) is carried out when the local temperature of heavy quark pair is below the dissociation temperature of quarkonium state $i$.
After that, whenever scattering happens to heavy quark or heavy antiquark, the Wigner production is updated~\cite{Song:2023ywt}, considering temperature-dependent quarkonium radius, which is obtained by solviing Schr\"{o}dinger equation with the heavy quark potential.

\begin{figure*}
\centering
\vspace*{1cm}       % Give the correct figure height in cm
\includegraphics[width=6.4cm,clip]{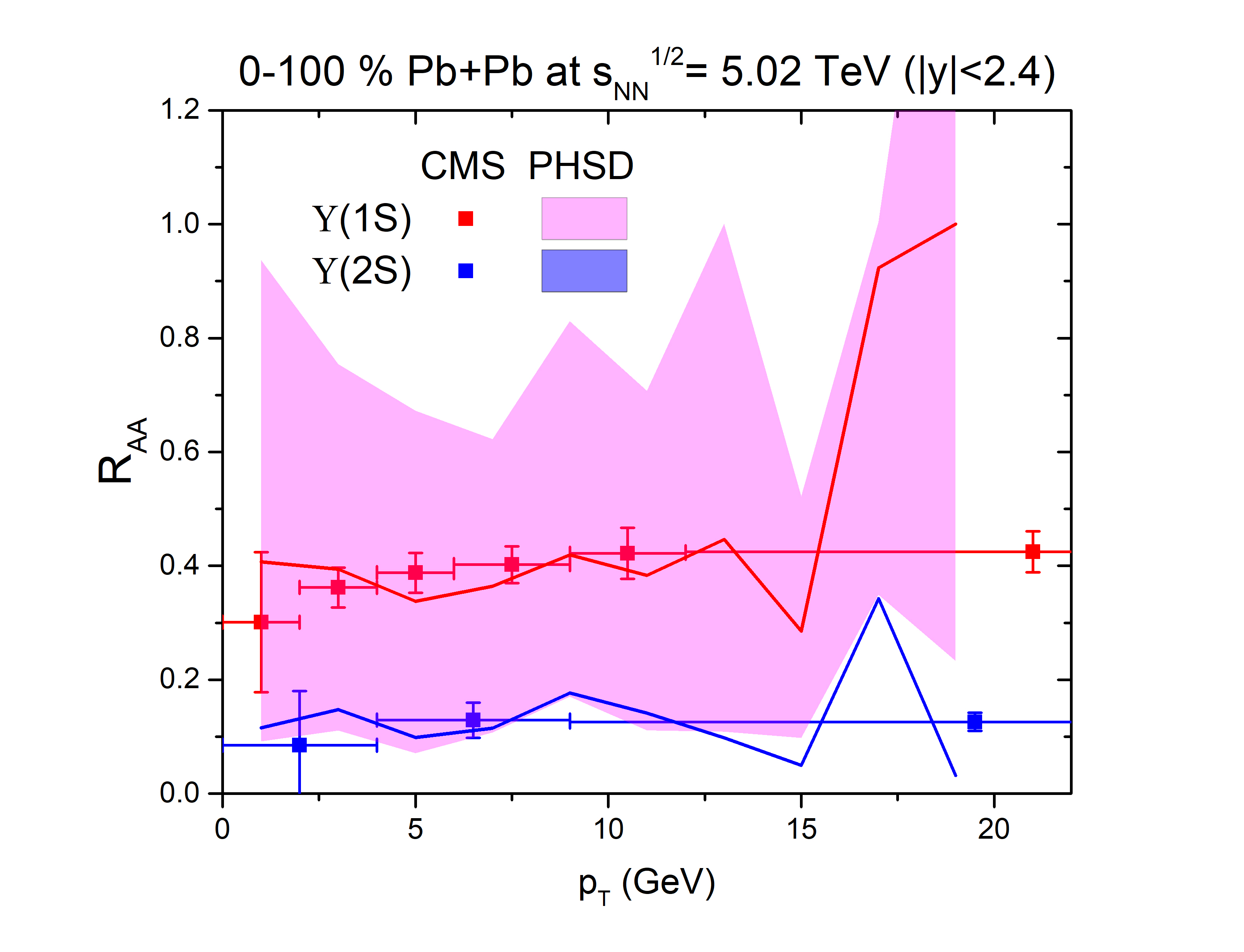}
\includegraphics[width=6.4cm,clip]{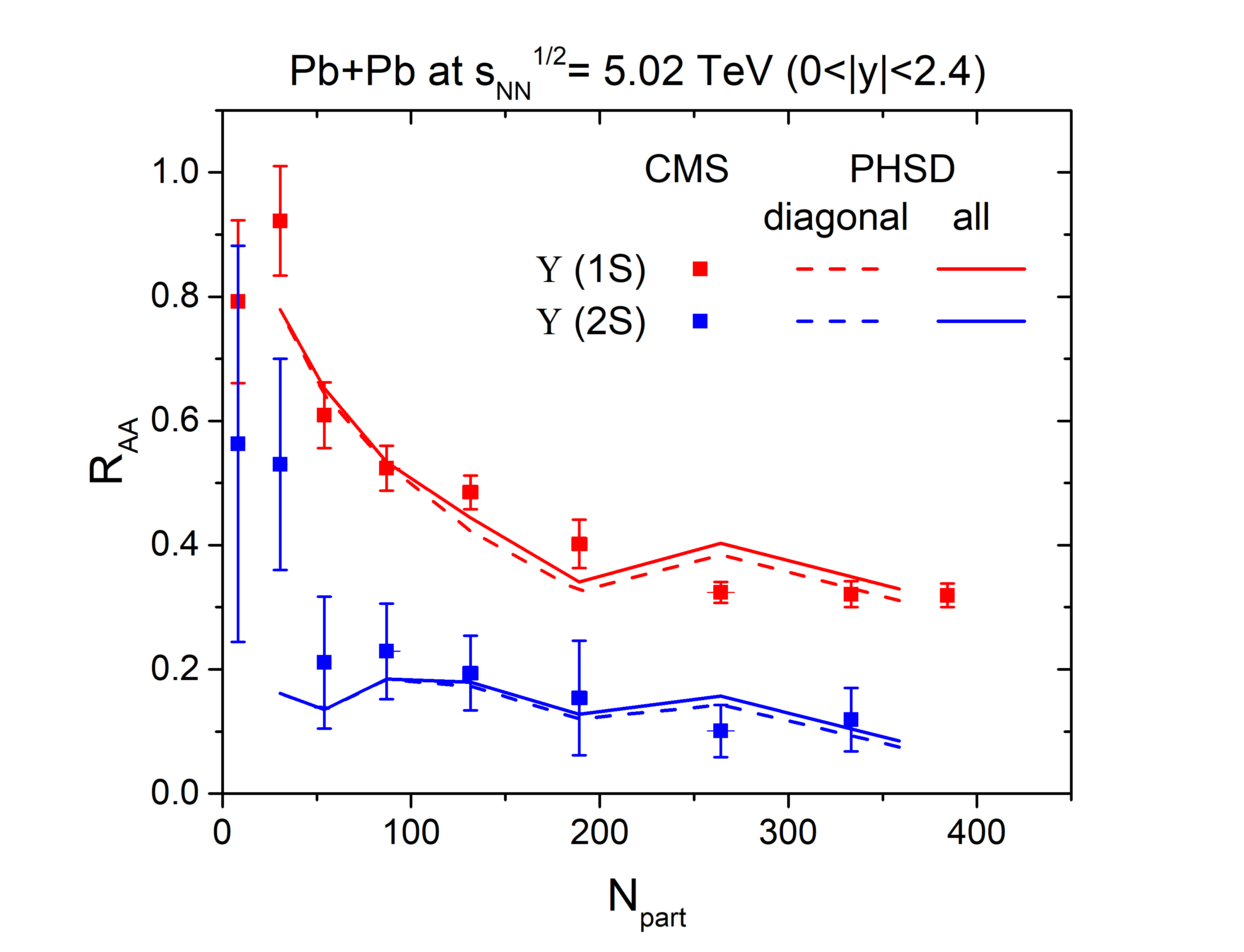}
\caption{$R_{AA}$ of $\Upsilon (1S)$ and $\Upsilon (1S)$ as a function of (left) transverse momentum and (right) number of participants in Pb+Pb collisions at $\sqrt{s}=5.02$ TeV in comparison with experimental data from the CMS Collaboration~\cite{CMS:2018zza}}
\label{AA}       % Give a unique label
\end{figure*}

Th left panel of Fig.~\ref{AA} displays $R_{AA}$ of $\Upsilon (1S)$ and $\Upsilon (1S)$ as a function of transverse momentum in Pb+Pb collisions at $\sqrt{s}=5.02$ TeV. 
The upper limit of the magenta band in the left panel indicates $R_{AA}$ of $\Upsilon (1S)$ at the dissociation temperature of $\Upsilon (1S)$ and the lower limit of the band $R_{AA}$ of $\Upsilon (1S)$ at $T_c$.
One can see that Wigner density of $\Upsilon (1S)$ decreases with time, because bottom and antibottom quarks are more and more separated from each other. 
The red solid line indicates $R_{AA}$ of $\Upsilon (1S)$, assuming only 10 \% of (anti)bottom quark scatterings update the Wigner density of $\Upsilon (1S)$, which is consistent with the experimental data from the CMS Collaboration~\cite{CMS:2018zza}.
In this case the interaction rate of $\Upsilon (1S)$ is roughly estimated to be 40-100 MeV at T=0.2-0.4 GeV~\cite{Song:2023zma,Andronic:2024oxz}.
As for $\Upsilon (2S)$ its dissociation temperature is close to $T_c$. That is why the blue band is so narrow that it looks like a line.
The right panel shows $R_{AA}$ as a function of the number of participants.
The solid lines include all contributions, while the dashed lines only diagonal contribution which originates from initial bottom quark pairs.
One can see that the off-diagonal contribution which originate from two different initial bottom quark pairs is little even in central Pb+Pb collisions  at $\sqrt{s}=5.02$ TeV, because the number of produced bottom quark pairs is not so large.

\section{Conclusion}
\label{sec-conclusion}
Normally quarkonium production in heavy-ion collisions is separately described from the production in pp collisions, though the nuclear modification factor ($R_{AA}$) is defined as their ratio. 
In this study we have used an unified method, the Remler's formalism, for quarkonium production in pp as well as heavy-ion collisions. 
Assuming that the interaction rate of bottom (anti)quark bound in quarkonium is 10 \% that of unbound bottom (anti)quark, our results well describe the experimental data in Pb+Pb collisions at $\sqrt{s}=5.02$ TeV.
Furthemore, we have found that the off-diagonal contribution is little even in LHC.

\section*{Acknowledgements}
%=================================================================================

We acknowledge support by the Deutsche Forschungsgemeinschaft (DFG, German Research Foundation) through the grant CRC-TR 211 'Strong-interaction matter under extreme conditions' - Project number 315477589 - TRR 211. This work is also supported by the European Union’s Horizon 2020 research and innovation program under grant agreement No 824093 (STRONG-2020). The computational resources have been provided by the LOEWE-Center for Scientific Computing and the "Green Cube" at GSI, Darmstadt, and by the Center for Scientific Computing (CSC) of the Goethe University, Frankfurt.

%
% BibTeX or Biber users please use (the style is already called in the class, ensure that the "woc.bst" style is in your local directory)
% \bibliography{your_bib_file} % Replace "your_bib_file" with the actual name of your .bib file

\end{document}